\documentclass[12pt]{article}%
\usepackage{graphicx}%
\usepackage[normalem]{ulem}
\usepackage{amsmath,amsthm,amsfonts,
amsbsy,amssymb,upref,enumerate,bigstrut,
color,mathtools,mathrsfs,float,bm,dsfont,scalefnt,mathtools}
\usepackage{natbib}
\usepackage{lipsum}
\usepackage{lmodern}
\usepackage{stmaryrd}
\usepackage{authblk} 
\usepackage{subfigure}
\usepackage{pgfplots}
\usetikzlibrary{pgfplots.fillbetween}

\usepackage{caption}
\usepackage{tikz}
\usetikzlibrary{intersections}

\usepackage[colorlinks=true,linkcolor=blue,citecolor=blue]{hyperref}

\usepackage[left=1in,top=1.2in,right=0.7in]{geometry}

\usepackage{booktabs}

\numberwithin{equation}{section} 

\makeatletter
\def\@seccntformat#1{\@ifundefined{#1@cntformat}%
	{\csname the#1\endcsname\quad}
	{\csname #1@cntformat\endcsname}
}
\makeatother

\newif\ifShowComments
\ShowCommentstrue
\def\strutdepth{\dp\strutbox}
\def\druk#1{\strut\vadjust{\kern-\strutdepth
        {\vtop to \strutdepth{%
                \baselineskip\strutdepth\vss
                        \llap{\hbox{#1}\quad}\null}}}}

\title{\bf
Length-biased Birnbaum-Saunders quantile regression with application to water evaporation
}

\author[1,2]{Helton Saulo\thanks{Corresponding author: heltonsaulo@gmail.com}}
\author[1]{Tailine Nonato}
\author[1]{Roberto Vila}
\affil[1]{Department of Statistics, University of
	 Bras\'ilia, Bras\'ilia, Brazil}
\affil[2]{
Department of Economics, Federal University of Pelotas, Pelotas, Brazil}

\begin{document}
\maketitle

\begin{abstract}
Length-biased distributions arise naturally in environmental, reliability, and economic studies where the sampling mechanism favors larger observational units. In this paper, we propose a quantile regression model based on the length-biased Birnbaum--Saunders (QLBS) distribution. The model is constructed through a reparameterization of the length-biased Birnbaum--Saunders distribution in terms of its quantile function, thereby allowing direct interpretation of covariate effects on conditional quantiles of the response variable. We derive the log-likelihood function and the corresponding score equations, and obtain maximum likelihood estimators via numerical optimization. Asymptotic and bootstrap confidence intervals are considered. Two types of residuals are proposed for model assessment, namely the generalized Cox--Snell and randomized quantile residuals. An elaborate Monte Carlo simulation study is carried out to evaluate the performance of the maximum likelihood estimators for several sample sizes and quantile levels. The proposed methodology is illustrated with a real meteorological data set from Brazil.
\end{abstract}



\section{Introduction}
\label{sec:1}
\noindent

Regression models play a central role in statistical analysis, providing a systematic framework for investigating the relationship between a response variable and a set of covariates. Among the families of regression models designed for positive and right-skewed response variables, those based on the Birnbaum--Saunders (BS) distribution have received considerable attention in the literature; see the review by \cite{bk:19}. \cite{ddlms:20} performed a comparative study of three regression approaches for asymmetric BS data, covering the formulations of \cite{rn:91}, \cite{lscb:14}, and \cite{bz:15}. It is relevant to note that the classical BS regression model expresses covariate effects through the conditional scale parameter, which is also the median in the BS case. Nevertheless, it may be inadequate when other quantiles are of interest.

Quantile regression, introduced by \cite{kb:78}, addresses this limitation by directly modeling conditional quantiles, thereby providing a more comprehensive characterization of the conditional distribution of the response variable. Within the BS framework, \cite{slgs:2020a,slgs:2020b} and \cite{lsgs:2020c} proposed quantile regression models based on the BS distribution, whereas \cite{ssls:20} developed log-symmetric quantile regression models that encompass the BS case as a special instance. These contributions have demonstrated that quantile regression approaches grounded on BS-type distributions are particularly suited to positive and skewed data commonly arising in environmental and economic contexts.

Weighted distributions, introduced by \cite{sa:01}, constitute a class of probability models that adjust the sampling probabilities of a given distribution to account for the mechanism by which observations are collected. \cite{p:06} argued that weighted distributions are especially relevant in environmental sciences, where data are often collected through nonexperimental or size-biased sampling schemes rather than simple random sampling. Length-biased distributions form an important subclass of weighted distributions, arising when the probability of selecting an observation is proportional to its size. In the context of the BS distribution, \cite{lsa:09} introduced the length-biased BS (LBS) distribution, deriving its moments and properties and illustrating its use with water quality data. Recently, \cite{ocsvils:23} proposed a regression model based on the LBS distribution and applied it to meteorological data. In that model, however, covariate effects are expressed through the scale parameter of the LBS distribution, which does not admit a direct quantile interpretation.

In this paper, we propose a quantile regression formulation based on the LBS distribution, hereafter referred to as the QLBS regression model. The model is formulated by reparameterizing the LBS distribution in terms of its quantile function, so that covariate effects are expressed directly in terms of conditional quantiles of the response. The secondary objectives of this paper are: (a)~to derive the log-likelihood function and the corresponding score equations, obtaining maximum likelihood estimators via numerical optimization; (b)~to establish asymptotic confidence intervals for the model parameters; (c)~to propose and evaluate diagnostic residuals for model assessment; (d)~to carry out an elaborate Monte Carlo simulation study to evaluate the performance of the proposed estimators for several sample sizes and quantile levels; and (e)~to illustrate the proposed model with a real data application to meteorological data.

The paper is organized as follows. In Section~\ref{sec:2}, we describe the LBS distribution and obtain its reparameterization in terms of quantiles. Section~\ref{sec:3} presents the QLBS regression model, including inference procedures, initial values, confidence intervals, and residual analysis. In Section~\ref{sec:4}, we carry out Monte Carlo simulations to evaluate the finite sample performance of the proposed estimators. An application to meteorological data from Brazil is presented in Section~\ref{sec:5}. Finally, Section~\ref{sec:6} contains some concluding remarks.

\section{The LBS distribution and its reparametrization by quantile}\label{sec:2}

In this section, we briefly describes the LBS distribution. Then, a novel parametrization based on the quantile parameter is obtained.

\subsection{Length-biased Birnbaum--Saunders distribution}\label{sec:2.1}

Let $Y$ be a nonnegative random variable with probability density function (PDF) $f_Y(\cdot)$ and finite mean $\mathrm{E}(Y)$. The length-biased version of $Y$, denoted by $T$, is defined through the PDF
\begin{equation}\label{LB-density}
f_T(t) = \frac{t\,f_Y(t)}{\mathrm{E}(Y)}, \qquad t>0,
\end{equation}
which arises naturally in situations where the sampling mechanism favors larger realizations of the underlying variable. When $Y$ follows a Birnbaum--Saunders distribution with shape parameter $\alpha>0$ and scale parameter $\theta>0$, denoted by $Y \sim \mathrm{BS}(\alpha,\theta)$, its mean is given by $\mathrm{E}(Y)=\theta(\alpha^2+2)/2$. Substituting this expression into \eqref{LB-density}, the resulting random variable $T$ follows a length-biased Birnbaum--Saunders (LBS) distribution, proposed by \cite{lsa:09}, whose PDF can be written as
\begin{equation}\label{eq:fdp}
f_T(t)
=
\phi(a_t)\,
\frac{1}{\alpha^{3}+2\alpha}
\left[
\left(\frac{t}{\theta}\right)^{1/2}
+
\left(\frac{\theta}{t}\right)^{1/2}
\right],
\quad t>0,
\end{equation}
and the cumulative distribution function (CDF) as
\begin{equation}\label{eq:cdf_LBS}
F_T(t)
=
\Phi(a_t)
+
\frac{\alpha^2}{2+\alpha^2}
\left[
\exp\!\left(\frac{2}{\alpha^2}\right)
\left\{
\Phi\!\left(\frac{\sqrt{4+\alpha^2 a_t^2}}{\alpha}\right)
-
1
\right\}
-
\phi(a_t)
\left(
a_t
+
\frac{\sqrt{4+\alpha^2 a_t^2}}{\alpha}
\right)
\right],
\qquad t>0,
\end{equation}
with
\begin{equation}
a_t = a_t(\alpha,\theta)
=
\frac{1}{\alpha}
\left[
\left(\frac{t}{\theta}\right)^{1/2}
-
\left(\frac{\theta}{t}\right)^{1/2}
\right]
\end{equation}
for $t>0$, $\alpha>0$ and $\theta>0$. In this case, we write $T \sim \mathrm{LBS}(\alpha,\theta)$.

In the LBS model, the parameter $\theta$ acts as a scale parameter, whereas $\alpha$ governs the degree of asymmetry and tail behavior of the distribution. Several properties of the $\mathrm{LBS}(\alpha,\theta)$ distribution are available in the literature and are summarized below. Let $T \sim \mathrm{LBS}(\alpha,\theta)$. Then \citep{lsa:09}:
\begin{enumerate}[(P1)]
\item for any constant $c>0$, the scaled variable $cT$ satisfies $cT \sim \mathrm{LBS}(\alpha,c\theta)$;
\item the mean of $T$ is
\[
\mathrm{E}(T)
=
\theta
\left(
\frac{2+4\alpha^2+3\alpha^4}{2+\alpha^2}
\right);
\]
\item the variance of $T$ is given by
\[
\mathrm{Var}(T)
=
\theta^2\alpha^2
\left[
\frac{4+17\alpha^2+24\alpha^4+6\alpha^6}{(2+\alpha^2)^2}
\right];
\]
\item for $r>-1$,
\[
\mathrm{E}\!\left(T^{-(r+1)}\right)
=
\frac{\mathrm{E}(Y^r)}{\theta^{2r}\mathrm{E}(Y)},
\qquad
Y \sim \mathrm{BS}(\alpha,\theta);
\]
\item the transformation
\[
U=\frac{1}{\alpha^2}
\left(
\frac{T}{\theta}
+
\frac{\theta}{T}
-
2
\right)
\]
follows a finite mixture of gamma distributions with PDF
\begin{equation}\label{eq:U_pdf_mixture}
f_U(u)
=
\pi f_{U_1}(u)
+
(1-\pi)f_{U_2}(u),
\end{equation}
where $\pi=2/(\alpha^2+2)$, $U_1 \sim \mathrm{Gamma}(1/2,2)$ and $U_2 \sim \mathrm{Gamma}(3/2,2)$.
\end{enumerate}

\subsection{A LBS distribution parametrized by its quantile}\label{sec:2.2}

Starting from Property (P5) of Section~\ref{sec:2.1}:
\begin{equation}\label{eq:U_def}
U=\frac{1}{\alpha^{2}}\left(\frac{T}{\theta}+\frac{\theta}{T}-2\right),
\end{equation}
we obtain the following stochastic relation
\begin{equation}\label{eq:T_isolated_final}
T
=
\frac{\theta}{4}
\left(
\alpha\sqrt{U}+\sqrt{\alpha^2U+4}
\right)^2.
\end{equation}
%
%
Given $\tau \in (0,1)$ and based on \eqref{eq:T_isolated_final}, note that the $\tau\times 100$th quantile of the LBS distribution is represented as
\begin{equation}\label{quantileBS}
Q_{\tau}=
\frac{\theta}{4}
\left(
\alpha\sqrt{u_{\tau}}+\sqrt{\alpha^2u_{\tau}+4}
\right)^2
\end{equation}
where $u_{\tau}$ is the $\tau\times 100$th quantile of a finite mixture of gamma distributions with PDF given in \eqref{eq:U_pdf_mixture}.

Let $\tau \in (0,1)$ be fixed and consider the transformation
\[
(\alpha,\theta) \mapsto (\alpha,Q_\tau),
\]
where $Q_\tau$ is the $\tau\times 100$th quantile of the $\mathrm{LBS}(\alpha,\theta)$ distribution defined in
\eqref{quantileBS}. For fixed $\alpha$ and $\tau$, the mapping between $\theta$ and $Q_\tau$ is one-to-one, so that the
LBS distribution can be reparameterized in terms of $(\alpha,Q_\tau)$. From \eqref{quantileBS}, the scale parameter $\theta$ can be expressed as
\begin{equation}\label{eq:theta_quantile}
\theta
=
\frac{4 Q_{\tau}}
{\kappa_{\tau}(\alpha)},
\end{equation}
where
\[
\kappa_{\tau}(\alpha)
=
\left(
\alpha\sqrt{u_{\tau}}+\sqrt{\alpha^2u_{\tau}+4}
\right)^2
\]
and $u_{\tau}$ denotes the $\tau\times 100$th quantile of the finite mixture of gamma distributions defined in
\eqref{eq:U_pdf_mixture}. Substituting \eqref{eq:theta_quantile} into the expressions of the PDF and CDF given in
\eqref{eq:fdp} and \eqref{eq:cdf_LBS}, respectively, yields the quantile-parameterized form of the LBS distribution.
Specifically, the probability density function of $T$ can be written as
\begin{equation}\label{eq:pdf_LBS_quantile}
f_T(t)
=
\phi(a_{t,\tau})\,
\frac{1}{\alpha^{3}+2\alpha}
\left[
\left(\frac{t\,\kappa_{\tau}(\alpha)}{4Q_\tau}\right)^{1/2}
+
\left(\frac{4Q_\tau}{t\,\kappa_{\tau}(\alpha)}\right)^{1/2}
\right],
\qquad t>0,
\end{equation}
where
\begin{equation}\label{eq:a_t_quantile}
a_{t,\tau}
=
\frac{1}{\alpha}
\left[
\left(\frac{t\,\kappa_{\tau}(\alpha)}{4Q_\tau}\right)^{1/2}
-
\left(\frac{4Q_\tau}{t\,\kappa_{\tau}(\alpha)}\right)^{1/2}
\right].
\end{equation}

Similarly, the cumulative distribution function of $T$ under this parameterization is given by
\begin{equation}\label{eq:cdf_LBS_quantile}
\begin{aligned}
F_T(t)
&=
\Phi(a_{t,\tau})
+
\frac{\alpha^2}{2+\alpha^2}
\Bigg[
\exp\!\left(\frac{2}{\alpha^2}\right)
\left\{
\Phi\!\left(
\frac{\sqrt{4+\alpha^2 a_{t,\tau}^2}}{\alpha}
\right)
-
1
\right\} \\
&\hspace{2.5cm}
-
\phi(a_{t,\tau})
\left(
a_{t,\tau}
+
\frac{\sqrt{4+\alpha^2 a_{t,\tau}^2}}{\alpha}
\right)
\Bigg],
\qquad t>0.
\end{aligned}
\end{equation}
Under this parameterization, we write
$
T \sim \mathrm{QLBS}(\alpha,Q_\tau),
$
emphasizing that $Q_\tau$ represents the $\tau$th quantile of the distribution. This formulation allows direct modeling
and interpretation of conditional quantiles, which is particularly appealing in regression contexts where the interest
lies in different parts of the response distribution rather than in its mean. Finally, the mean and variance of $T$ under the quantile parameterization can be obtained by substituting
\eqref{eq:theta_quantile} into the expressions in Properties (P2) and (P3), yielding
\[
\mathrm{E}(T)
=
\frac{2 Q_\tau}{\kappa_{\tau}(\alpha)}
\left(
\frac{2+4\alpha^2+3\alpha^4}{2+\alpha^2}
\right),
\qquad
\mathrm{Var}(T)
=
\left(
\frac{2 Q_\tau}{\kappa_{\tau}(\alpha)}
\right)^2
\alpha^2
\left[
\frac{4+17\alpha^2+24\alpha^4+6\alpha^6}{(2+\alpha^2)^2}
\right].
\]

\section{The QLBS regression model}\label{sec:3}

Let $T_1, T_2,\ldots, T_n$ denote independent random variables, where
\[
T_i \sim \mathrm{QLBS}(\alpha_i, Q_{\tau,i}), \quad i=1,\ldots,n,
\]
with observed values denoted by $t_1, t_2, \ldots, t_n$, respectively, and where
$Q_{\tau,i}$ represents the $\tau\times 100$th conditional quantile of $T_i$. The QLBS regression model is formulated as
\begin{equation}\label{eq:pred_quantile}
\begin{aligned}
\eta_{1i}
=
g_1(Q_{\tau,i})
=
\sum_{j=1}^{p} x_{ij}\beta_j
=
\bm{x}_i^{\top}\bm{\beta},
\quad i=1,\ldots,n,
\\
\eta_{2i}
=
g_2(\alpha_i)
=
\sum_{j=1}^{q} w_{ij}\rho_j
=
\bm{w}_i^{\top}\bm{\rho},
\quad i=1,\ldots,n,
\end{aligned}
\end{equation}
where $\bm{\beta}=(\beta_1,\ldots,\beta_p)^{\top}$ and
$\bm{\rho}=(\rho_1,\ldots,\rho_q)^{\top}$ are vectors of unknown regression
parameters, with $\bm{\beta}\in\mathbb{R}^p$, $\bm{\rho}\in\mathbb{R}^q$ and
$p+q<n$. The covariate vectors
$\bm{x}_i^{\top}=(x_{i1},\ldots,x_{ip})$ and
$\bm{w}_i^{\top}=(w_{i1},\ldots,w_{iq})$ correspond to the $i$th rows of the
design matrices $\mathbf{X}\in\mathbb{R}^{n\times p}$ and
$\mathbf{W}\in\mathbb{R}^{n\times q}$, respectively, which are assumed to be of
full rank, that is, $\mathrm{rank}(\mathbf{X})=p$ and
$\mathrm{rank}(\mathbf{W})=q$. Typically, the first columns of $\mathbf{X}$ and
$\mathbf{W}$ are vectors of ones, denoted by
$\mathbf{x}_{\cdot 1}=\bm{1}_n^{\top}$ and
$\mathbf{w}_{\cdot 1}=\bm{1}_n^{\top}$, where $\bm{1}_n$ is an $n$-dimensional
vector of ones.  The link functions $g_1:(0,\infty)\to\mathbb{R}$ and
$g_2:(0,\infty)\to\mathbb{R}$ are assumed to be invertible and at least twice
continuously differentiable. Common choices for $g_1(\cdot)$ include the log
link, $g_1(u)=\log(u)$, and the square-root link, $g_1(u)=\sqrt{u}$. Analogous choices may
be adopted for $g_2(\cdot)$ when modeling the shape parameter $\alpha_i$.

\subsection{Parameter estimation}\label{sec:3.1}
For notational convenience, define

\begin{equation}\label{eq:AandB}
A = \left(\frac{t \kappa_{\tau}(\alpha)}{4Q_\tau}\right)^{1/2} \hspace{10px} \text{and} \hspace{10px}
B = \left(\frac{4Q_\tau}{t \kappa_{\tau}(\alpha)}\right)^{1/2},
\end{equation}

so that $a_{t,\tau} = \frac{1}{\alpha}\left[A - B\right]$ and consequently $a_{t,\tau}^2 = \frac{1}{\alpha^2}\left[A-B\right]^2$.

Substituting (\ref{eq:AandB}) into the probability density function of $T$ (\ref{eq:pdf_LBS_quantile}) the PDF is given by

\begin{equation}\label{eq:PDF_AandB}
f_T(t) = \phi(a_{t,\tau})  \frac{1}{\alpha^3 + 2\alpha} (A + B) = \frac{1}{\sqrt{2\pi}}  \exp\left(-\frac{a_{t,\tau}^2}{2}\right)  \frac{1}{\alpha^3 + 2\alpha} (A + B).
\end{equation} 

The corresponding likelihood function is

\begin{equation}\label{eq:likelihood}
\begin{aligned}
L(\alpha, Q_\tau) 
&= \prod_{i=1}^n f_T(t_i) \\
&= \frac{1}{(2\pi)^{n/2}}  \exp\left(-\frac{1}{2} \sum_{i=1}^n a_{t_i,\tau}^2\right)  \frac{1}{(\alpha^3 + 2 \alpha)^n} \prod_{i=1}^n \left(A_i + B_i\right).
\end{aligned}
\end{equation}

Then, the log-likelihood function can be expressed as

\begin{equation}\label{eq:loglik}
\begin{aligned}
\ell(\alpha, Q_\tau) = -\frac{n}{2} \log(2\pi) - n \log(\alpha^3 + 2\alpha) - \frac{1}{2} \sum_{i=1}^n a_{t_i,\tau}^2 + \sum_{i=1}^n \log(A_i + B_i).
\end{aligned}
\end{equation}

The maximum likelihood estimators of the QLBS regression model parameters are the solutions of the system $\dot{\ell}(\bm{\delta}) = \bm{0}$, where $\bm{\delta} = (\bm{\beta}^{\top}, \bm{\rho}^{\top})^{\top}$ and the gradient vector takes the form
\begin{equation}\label{eq:gradient}
\dot{\ell}(\bm{\delta})
= \begin{pmatrix} \partial\ell/\partial\bm{\beta} \\ \partial\ell/\partial\bm{\rho} \end{pmatrix}
= \begin{pmatrix} \mathbf{X}^{\top}\mathbf{A}\bm{z} \\ \mathbf{W}^{\top}\mathbf{B}\bm{c} \end{pmatrix},
\end{equation}
where $\mathbf{A} = \mathrm{diag}(a_1,\ldots,a_n)$, $\mathbf{B} = \mathrm{diag}(b_1,\ldots,b_n)$, $\bm{z} = (z_1,\ldots,z_n)^{\top}$ and $\bm{c} = (c_1,\ldots,c_n)^{\top}$, with $a_i = [g_1'(Q_{\tau,i})]^{-1}$ and $b_i = [g_2'(\alpha_i)]^{-1}$ being the inverses of the link function derivatives. The vectors $\bm{z}$ and $\bm{c}$ contain the partial derivatives of the individual log-likelihood contributions with respect to $Q_{\tau,i}$ and $\alpha_i$, respectively, obtained by applying the chain rule
\begin{equation}\label{eq:chain_rule}
\frac{\partial\ell}{\partial\bm{\beta}} = \sum_{i=1}^{n} z_i\, a_i\, \bm{x}_i, \qquad
\frac{\partial\ell}{\partial\bm{\rho}} = \sum_{i=1}^{n} c_i\, b_i\, \bm{w}_i.
\end{equation}
The scalar components $z_i$ and $c_i$ are obtained by differentiating (\ref{eq:loglik}) with respect to $Q_{\tau,i}$ and $\alpha_i$, yielding
\begin{equation}\label{eq:score_z}
z_i
= \frac{\partial \ell_i}{\partial Q_{\tau,i}}
= \frac{A_i^2 - B_i^2}{2\alpha_i^2\, Q_{\tau,i}}
  - \frac{A_i - B_i}{2\,Q_{\tau,i}(A_i + B_i)},
\end{equation}
\begin{equation}\label{eq:score_c}
c_i
= \frac{\partial \ell_i}{\partial \alpha_i}
= -\frac{3\alpha_i^2 + 2}{\alpha_i^3 + 2\alpha_i}
  + \frac{(A_i - B_i)^2}{\alpha_i^3}
  - \frac{m_i(A_i^2 - B_i^2)}{2\alpha_i^2}
  + \frac{m_i(A_i - B_i)}{2(A_i + B_i)},
\end{equation}
where $m_i = \kappa_{\tau}'(\alpha_i)/\kappa_{\tau}(\alpha_i)$ is the logarithmic derivative of $\kappa_{\tau}(\cdot)$ evaluated at $\alpha_i$, and $A_i$, $B_i$ are as defined in (\ref{eq:AandB}) with $(\alpha, Q_{\tau})$ replaced by $(\alpha_i, Q_{\tau,i})$. These equations do not admit closed-form solutions, and the estimators are therefore obtained via iterative numerical methods; see \cite{mjm:00}.

The observed information matrix is obtained from the Hessian of the log-likelihood function evaluated at $\widehat{\bm{\delta}}$, which takes the form
\begin{equation}\label{eq:hessian}
\ddot{\ell}(\bm{\delta}) =
\begin{pmatrix}
\ddot{\ell}_{\bm{\beta\beta}} & \ddot{\ell}_{\bm{\beta\rho}} \\
\ddot{\ell}_{\bm{\rho\beta}} & \ddot{\ell}_{\bm{\rho\rho}}
\end{pmatrix}
=
\begin{pmatrix}
\mathbf{X}^{\top}\mathbf{V}\mathbf{X} & \mathbf{X}^{\top}\mathbf{H}\mathbf{W} \\
\mathbf{W}^{\top}\mathbf{H}^{\top}\mathbf{X} & \mathbf{W}^{\top}\mathbf{U}\mathbf{W}
\end{pmatrix},
\end{equation}
where $\mathbf{V} = \mathbf{Z}'\mathbf{A}^2 + \mathbf{Z}\mathbf{D}\mathbf{A}$, $\mathbf{H} = \mathbf{K}\mathbf{B}\mathbf{A}$, and $\mathbf{U} = \mathbf{C}'\mathbf{B}^2 + \mathbf{C}\mathbf{E}\mathbf{B}$ are $n\times n$ diagonal matrices, with
\begin{equation}\label{eq:hessian_diags}
\begin{aligned}
v_i &= z_i'\, a_i^2 + z_i\, d_i\, a_i, \\
h_i &= k_i\, b_i\, a_i, \\
u_i &= c_i'\, b_i^2 + c_i\, e_i\, b_i,
\end{aligned}
\end{equation}
and $\mathbf{Z}' = \mathrm{diag}(z_1',\ldots,z_n')$, $\mathbf{C}' = \mathrm{diag}(c_1',\ldots,c_n')$, $\mathbf{K} = \mathrm{diag}(k_1,\ldots,k_n)$, $\mathbf{D} = \mathrm{diag}(d_1,\ldots,d_n)$, $\mathbf{E} = \mathrm{diag}(e_1,\ldots,e_n)$. Here $z_i' = \partial^2\ell_i/\partial Q_{\tau,i}^2$, $c_i' = \partial^2\ell_i/\partial\alpha_i^2$, $k_i = \partial^2\ell_i/(\partial Q_{\tau,i}\,\partial\alpha_i)$, $d_i = -g_1''(Q_{\tau,i})/[g_1'(Q_{\tau,i})]^2$, and $e_i = -g_2''(\alpha_i)/[g_2'(\alpha_i)]^2$.
The explicit expressions for $z_i'$, $c_i'$, and $k_i$ are given by
\begin{equation}\label{eq:hessian_z2}
z_i'
= \frac{1}{Q_{\tau,i}^2}\left[
  -\frac{A_i^2}{\alpha_i^2}
  + \frac{1}{(A_i+B_i)^2}
  + \frac{A_i - B_i}{2(A_i+B_i)}
\right],
\end{equation}
\begin{equation}\label{eq:hessian_k}
k_i
= \frac{1}{Q_{\tau,i}}\left[
  \frac{m_i(A_i^2+B_i^2)}{2\alpha_i^2}
  - \frac{A_i^2-B_i^2}{\alpha_i^3}
  - \frac{m_i}{(A_i+B_i)^2}
\right],
\end{equation}
and $c_i'$ is obtained by differentiating (\ref{eq:score_c}) with respect to $\alpha_i$, which involves the derivative $m_i' = \mathrm{d}m_i/\mathrm{d}\alpha_i$ and is computed numerically in practice.

\subsection{Initial values}\label{sec:3.2}

In order to initiate the numerical optimization of the log-likelihood function, initial values for the parameters are required. The initial value for $\bm{\beta}=(\beta_1,\ldots,\beta_p)^\top$ can be obtained by regressing $g_1(t_i)$ on the covariate vector $\bm{x}_i$ using ordinary least squares, yielding
\[
\bm{\beta}_0 = (\mathbf{X}^\top\mathbf{X})^{-1}\mathbf{X}^\top g_1(\mathbf{t}),
\]
where $\mathbf{t}=(t_1,\ldots,t_n)^\top$ and $g_1$ is the link functon as in \eqref{eq:pred_quantile}. Given this initial estimate, define $\widehat{\theta}_{i,0}=g_1^{-1}(\bm{x}_i^\top\bm{\beta}_0)$ and let $\widehat{\alpha}_{i,0}=\sqrt{(t_i/\widehat{\theta}_{i,0}+\widehat{\theta}_{i,0}/t_i-2}$ for $i=1,\ldots,n$. The initial value for $\bm{\rho}=(\rho_1,\ldots,\rho_q)^\top$ is then obtained by regressing $g_2(\widehat{\alpha}_{i,0})$ on $\bm{w}_i$ via ordinary least squares, yielding
\[
\bm{\rho}_0 = (\mathbf{W}^\top\mathbf{W})^{-1}\mathbf{W}^\top g_2\!\left(\widehat{\bm{\alpha}}_0\right),
\]
where $\widehat{\bm{\alpha}}_0=(\widehat{\alpha}_{1,0},\ldots,\widehat{\alpha}_{n,0})^\top$ and $g_2$ is the link functon as in \eqref{eq:pred_quantile}. This procedure follows the approach proposed by \cite{ocsvils:23} for the LBS regression model. The initial vector $\bm{\delta}_0=(\bm{\beta}_0^\top,\bm{\rho}_0^\top)^\top$ is then used as the starting point in the numerical maximization of the log-likelihood function.

\subsubsection{Confidence intervals}\label{sec:3.3}

In this subsection, we derive confidence intervals (CIs) for the model parameters using the asymptotic properties of maximum likelihood estimators.

Under standard regularity conditions \citep{ch:74}, the maximum likelihood estimator $\widehat{\bm{\delta}}$ satisfies the asymptotic normality property
\begin{equation}\label{eq:asym_normal}
\sqrt{n}(\widehat{\bm{\delta}}-\bm{\delta})\xrightarrow{d}
\mathrm{N}_{p+q}\!\left(\bm{0}_{p+q},\,\Sigma_{\bm{\delta}}\right),
\end{equation}
where $\xrightarrow{d}$ denotes convergence in distribution and $\Sigma_{\bm{\delta}}$ is the asymptotic covariance matrix of $\widehat{\bm{\delta}}$, which is approximated by the inverse of the observed Fisher information matrix, that is, $\Sigma_{\bm{\delta}}\approx[-\ddot{\ell}(\widehat{\bm{\delta}})]^{-1}$. Based on \eqref{eq:asym_normal}, the $100(1-\gamma)\%$ asymptotic CI (ACI) for the $j$th component $\delta_j$ of $\bm{\delta}$ takes the form
\begin{equation}\label{eq:ACI}
\left(\widehat{\delta}_j - z_{1-\gamma/2}\,\widehat{\Sigma}_{\delta_{jj}}^{1/2},\;
      \widehat{\delta}_j + z_{1-\gamma/2}\,\widehat{\Sigma}_{\delta_{jj}}^{1/2}\right),
\end{equation}
where $z_{1-\gamma/2}$ is the $(1-\gamma/2)$th quantile of the standard normal distribution, $\widehat{\Sigma}_{\delta_{jj}}$ is the $j$th diagonal element of $\widehat{\Sigma}_\delta\approx[-\ddot{\ell}(\widehat{\bm{\delta}})]^{-1}$, and $j=1,\ldots,p+q$.


\subsection{Residual analysis}\label{sec:3.4}

In order to assess the validity of the model assumptions and to evaluate the goodness of fit, we propose two types of residuals for the QLBS regression model. Let $\widehat{Q}_{\tau,i}=g_1^{-1}(\bm{x}_i^\top\widehat{\bm{\beta}})$ and $\widehat{\alpha}_i=g_2^{-1}(\bm{w}_i^\top\widehat{\bm{\rho}})$ denote the fitted quantile and shape parameter for the $i$th observation, and let $\widehat{\theta}_i=4\widehat{Q}_{\tau,i}/\kappa_\tau(\widehat{\alpha}_i)$ be the derived scale parameter. The fitted survival function is expressed as $\widehat{S}_T(t_i)=1-F_T(t_i;\widehat{\alpha}_i,\widehat{\theta}_i)$, where $F_T(\cdot)$ is the CDF of the LBS distribution given in \eqref{eq:cdf_LBS}.

The first residual is the generalized Cox--Snell (GCS) residual \citep{ocsvils:23}, given by
\begin{equation}\label{eq:r_GCS}
r_i^{\mathrm{GCS}} = -\log\!\left(\widehat{S}_T(t_i)\right), \quad i=1,\ldots,n.
\end{equation}
As mentioned in \cite{ssls:20}, the GCS residual is asymptotically standard exponential, $\mathrm{Exp}(1)$, when the model is correctly specified.

The second residual is the randomized quantile (RQ) residual, given by
\begin{equation}\label{eq:r_RQ}
r_i^{\mathrm{RQ}} = \Phi^{-1}\!\left(\widehat{S}_T(t_i)\right), \quad i=1,\ldots,n,
\end{equation}
where $\Phi^{-1}(\cdot)$ is the inverse function of the standard normal CDF. The RQ residual follows a standard normal distribution when the model is correctly specified; see \cite{ssls:20}. In order to assess the goodness of fit of the residuals, simulation envelopes based on simulated samples from the reference distributions are constructed for each type of residual.

\section{Monte Carlo simulation study}\label{sec:4}

In this section, we carry out Monte Carlo simulation studies to evaluate the finite sample performance of the maximum likelihood estimators for the QLBS regression model. We also assess the performance of the CIs described in Section~\ref{sec:3.3} as well as the $r^{\mathrm{GCS}}$ and $r^{\mathrm{RQ}}$ residuals discussed in Section~\ref{sec:3.4}. We use the R software \citep{rmanual:20} for all numerical calculations. The simulation scenario considers sample sizes $n \in \{50, 100, 200, 400\}$ and quantile levels $\tau \in \{0.25, 0.50, 0.75\}$. For each combination, $B = 1{,}000$ Monte Carlo replications are carried out.

The data are generated from the QLBS regression model with one continuous covariate in each sub-model. Specifically, for each replication, the covariates $x_{1i}$ and $w_{1i}$ are drawn independently from a uniform distribution on $(-1, 1)$, and the response $t_i$ is generated from $T_i \sim \mathrm{QLBS}(\alpha_i, Q_{\tau,i})$, $i = 1, \ldots, n$, where the conditional quantile and shape parameter are specified through the log-link models
\begin{equation}\label{eq:sim_model}
\log(Q_{\tau,i}) = \beta_0 + \beta_1 x_{1i}, \qquad
\log(\alpha_i) = \rho_0 + \rho_1 w_{1i},
\end{equation}
with true parameter values $\bm{\delta}^* = (\beta_0^*, \beta_1^*, \rho_0^*, \rho_1^*)^{\top} = (1, -1, \log 0.25, 0.5)^{\top}$.


We assess the accuracy of the maximum likelihood estimators, we compute, for each parameter $\delta_j \in \{\beta_0, \beta_1, \rho_0, \rho_1\}$, the empirical mean, bias, and mean squared error (MSE), defined as
\begin{equation}\label{eq:sim_metrics}
\overline{\hat{\delta}}_j = \frac{1}{B}\sum_{b=1}^{B}\hat{\delta}_{j}^{(b)}, \quad
\mathrm{Bias}(\hat{\delta}_j) = \overline{\hat{\delta}}_j - \delta_j^*, \quad
\mathrm{MSE}(\hat{\delta}_j) = \frac{1}{B}\sum_{b=1}^{B}\bigl(\hat{\delta}_{j}^{(b)} - \delta_j^*\bigr)^2,
\end{equation}
where $\hat{\delta}_{j}^{(b)}$ denotes the estimate obtained in the $b$th replication. We evaluate the asymptotic confidence intervals described in Section~\ref{sec:3.3}, by reporting the empirical coverage probability (CP), defined as the proportion of replications for which the 95\% ACI contains the true value $\delta_j^*$, that is,
\begin{equation}\label{eq:sim_cp}
\mathrm{CP}(\hat{\delta}_j) = \frac{1}{B}\sum_{b=1}^{B} \mathbf{1}\!\left(\hat{\delta}_{j}^{(b)} + z_{0.025}\,\widehat{\mathrm{SE}}^{(b)}_{j} \;<\; \delta_j^* \;<\; \hat{\delta}_{j}^{(b)} + z_{0.975}\,\widehat{\mathrm{SE}}^{(b)}_{j}\right),
\end{equation}
where $z_p$ denotes the $p$th quantile of the standard normal distribution and $\widehat{\mathrm{SE}}^{(b)}_{j}$ is the standard error of $\hat{\delta}_j^{(b)}$, obtained from the observed information matrix as described in Section~\ref{sec:3.1}. Under correct model specification and for large $n$, the CP should approach the nominal level of 95\%.

\subsection{Maximum likelihood estimates}

In order to assess the performance of the maximum likelihood estimators, we report the empirical mean, bias, mean squared error (MSE), and coverage probability (CP) of the 95\% asymptotic confidence intervals (ACIs). Tables~\ref{tab:sim_tau025}--\ref{tab:sim_tau075} present the results for $\tau = 0.25$, $0.50$, and $0.75$, respectively. From Tables~\ref{tab:sim_tau025}--\ref{tab:sim_tau075}, we observe that, as the sample size increases, the empirical means tend to the respective true parameter values for all four parameters. Furthermore, the empirical bias and MSE both decrease as expected when the sample size increases, for both the quantile sub-model parameter estimators $(\widehat\beta_0, \widehat\beta_1)$ and the shape sub-model parameter estimators $(\widehat\rho_0, \widehat\rho_1)$. Regarding the coverage probabilities, the CP values are generally close to the nominal level of 95\% across all parameters and quantile levels. For small samples ($n = 50$), some undercoverage is observed. Nevertheless, as $n$ increases, the CP values approach and stabilize near the nominal 95\% level for all parameters and quantile levels. These results confirm the good performance of the maximum likelihood estimators and the asymptotic confidence intervals.

\begin{table}[!ht]
\centering
\caption{Empirical mean, bias, MSE, and coverage probability (CP, \%) of the MLE for the QLBS regression model with $\tau = 0.25$, $\beta_0 = 1$, $\beta_1 = -1$, $\rho_0 = \log(0.25)$, and $\rho_1 = 0.5$.}
\label{tab:sim_tau025}
\begin{tabular}{clrrrr}
\toprule
$n$ & Parameter & Mean & Bias & MSE & CP (\%) \\
\midrule
50 & $\hat{\beta}_0$ & 1.0005 & 0.0005 & 0.0013 & 93.60 \\
 & $\hat{\beta}_1$ & -0.9971 & 0.0029 & 0.0044 & 91.70 \\
 & $\hat{\rho}_0$ & -1.4344 & -0.0481 & 0.0135 & 91.30 \\
 & $\hat{\rho}_1$ & 0.5287 & 0.0287 & 0.0477 & 92.30 \\
\midrule
100 & $\hat{\beta}_0$ & 0.9997 & -0.0003 & 0.0006 & 93.50 \\
 & $\hat{\beta}_1$ & -1.0012 & -0.0012 & 0.0015 & 94.40 \\
 & $\hat{\rho}_0$ & -1.4066 & -0.0203 & 0.0056 & 93.40 \\
 & $\hat{\rho}_1$ & 0.5114 & 0.0114 & 0.0164 & 94.20 \\
\midrule
200 & $\hat{\beta}_0$ & 0.9993 & -0.0007 & 0.0003 & 94.40 \\
 & $\hat{\beta}_1$ & -1.0001 & -0.0001 & 0.0009 & 95.30 \\
 & $\hat{\rho}_0$ & -1.3957 & -0.0094 & 0.0026 & 94.90 \\
 & $\hat{\rho}_1$ & 0.5025 & 0.0025 & 0.0083 & 94.90 \\
\midrule
400 & $\hat{\beta}_0$ & 0.9999 & -0.0001 & 0.0001 & 95.10 \\
 & $\hat{\beta}_1$ & -0.9992 & 0.0008 & 0.0004 & 95.50 \\
 & $\hat{\rho}_0$ & -1.3909 & -0.0046 & 0.0013 & 95.30 \\
 & $\hat{\rho}_1$ & 0.4985 & -0.0015 & 0.0036 & 95.50 \\
\bottomrule
\end{tabular}
\end{table}

\begin{table}[!ht]
\centering
\caption{Empirical mean, bias, MSE, and coverage probability (CP, \%) of the MLE for the QLBS regression model with $\tau = 0.50$, $\beta_0 = 1$, $\beta_1 = -1$, $\rho_0 = \log(0.25)$, and $\rho_1 = 0.5$.}
\label{tab:sim_tau050}
\begin{tabular}{clrrrr}
\toprule
$n$ & Parameter & Mean & Bias & MSE & CP (\%) \\
\midrule
50 & $\hat{\beta}_0$ & 0.9978 & -0.0022 & 0.0011 & 94.20 \\
 & $\hat{\beta}_1$ & -0.9995 & 0.0005 & 0.0033 & 92.70 \\
 & $\hat{\rho}_0$ & -1.4259 & -0.0396 & 0.0130 & 90.60 \\
 & $\hat{\rho}_1$ & 0.5187 & 0.0187 & 0.0331 & 94.80 \\
\midrule
100 & $\hat{\beta}_0$ & 0.9997 & -0.0003 & 0.0006 & 94.30 \\
 & $\hat{\beta}_1$ & -1.0022 & -0.0022 & 0.0018 & 94.10 \\
 & $\hat{\rho}_0$ & -1.4056 & -0.0193 & 0.0056 & 93.70 \\
 & $\hat{\rho}_1$ & 0.5097 & 0.0097 & 0.0174 & 94.80 \\
\midrule
200 & $\hat{\beta}_0$ & 0.9994 & -0.0006 & 0.0003 & 93.40 \\
 & $\hat{\beta}_1$ & -0.9994 & 0.0006 & 0.0010 & 94.10 \\
 & $\hat{\rho}_0$ & -1.3950 & -0.0087 & 0.0025 & 94.60 \\
 & $\hat{\rho}_1$ & 0.5008 & 0.0008 & 0.0092 & 94.80 \\
\midrule
400 & $\hat{\beta}_0$ & 1.0004 & 0.0004 & 0.0001 & 94.60 \\
 & $\hat{\beta}_1$ & -1.0001 & -0.0001 & 0.0004 & 95.00 \\
 & $\hat{\rho}_0$ & -1.3904 & -0.0041 & 0.0012 & 95.60 \\
 & $\hat{\rho}_1$ & 0.4984 & -0.0016 & 0.0037 & 95.40 \\
\bottomrule
\end{tabular}
\end{table}

\begin{table}[H]
\centering
\caption{Empirical mean, bias, MSE, and coverage probability (CP, \%) of the MLE for the QLBS regression model with $\tau = 0.75$, $\beta_0 = 1$, $\beta_1 = -1$, $\rho_0 = \log(0.25)$, and $\rho_1 = 0.5$.}
\label{tab:sim_tau075}
\begin{tabular}{clrrrr}
\toprule
$n$ & Parameter & Mean & Bias & MSE & CP (\%) \\
\midrule
50 & $\hat{\beta}_0$ & 0.9953 & -0.0047 & 0.0015 & 92.20 \\
 & $\hat{\beta}_1$ & -1.0011 & -0.0011 & 0.0042 & 93.80 \\
 & $\hat{\rho}_0$ & -1.4225 & -0.0362 & 0.0120 & 92.50 \\
 & $\hat{\rho}_1$ & 0.5215 & 0.0215 & 0.0259 & 95.00 \\
\midrule
100 & $\hat{\beta}_0$ & 0.9966 & -0.0034 & 0.0007 & 93.10 \\
 & $\hat{\beta}_1$ & -0.9992 & 0.0008 & 0.0015 & 93.90 \\
 & $\hat{\rho}_0$ & -1.4047 & -0.0184 & 0.0056 & 94.10 \\
 & $\hat{\rho}_1$ & 0.5038 & 0.0038 & 0.0135 & 94.40 \\
\midrule
200 & $\hat{\beta}_0$ & 0.9974 & -0.0026 & 0.0004 & 94.10 \\
 & $\hat{\beta}_1$ & -1.0009 & -0.0009 & 0.0008 & 94.20 \\
 & $\hat{\rho}_0$ & -1.3977 & -0.0114 & 0.0028 & 94.40 \\
 & $\hat{\rho}_1$ & 0.5050 & 0.0050 & 0.0072 & 94.50 \\
\midrule
400 & $\hat{\beta}_0$ & 0.9988 & -0.0012 & 0.0002 & 94.10 \\
 & $\hat{\beta}_1$ & -1.0018 & -0.0018 & 0.0004 & 95.20 \\
 & $\hat{\rho}_0$ & -1.3924 & -0.0061 & 0.0014 & 93.90 \\
 & $\hat{\rho}_1$ & 0.5032 & 0.0032 & 0.0031 & 95.20 \\
\bottomrule
\end{tabular}
\end{table}

\subsection{Empirical distribution of residuals}

Table~\ref{tab:residuals_sim} presents the empirical mean, standard deviation (SD), coefficient of skewness (CS), and kurtosis (CK) of the $r^{\mathrm{GCS}}$ and $r^{\mathrm{RQ}}$ residuals for $\tau = 0.25$, $0.50$, and $0.75$. The reference values are: for $r^{\mathrm{GCS}}$, mean $= 1$, SD $= 1$, CS $= 2$, CK $= 9$; for $r^{\mathrm{RQ}}$, mean $= 0$, SD $= 1$, CS $= 0$, CK $= 3$. As mentioned in Section~\ref{sec:3.4}, these residuals conform well with the reference distributions when the model is correctly specified. From Table~\ref{tab:residuals_sim}, we note that the empirical moments of both residuals approach their reference values as the sample size increases across all considered quantile levels, which provides empirical evidence for the adequacy of the proposed residuals under the QLBS regression model.

\begin{table}[!ht]
\centering
\caption{Empirical mean, standard deviation (SD), coefficient of skewness (CS), and kurtosis (CK) of the $r^{\mathrm{GCS}}$ and $r^{\mathrm{RQ}}$ residuals for the QLBS regression model.}
\label{tab:residuals_sim}
\begin{tabular}{llrrrrrrrr}
\toprule
 & & \multicolumn{4}{c}{$r^{\mathrm{GCS}}$} & \multicolumn{4}{c}{$r^{\mathrm{RQ}}$} \\
\cmidrule(lr){3-6}\cmidrule(lr){7-10}
$\tau$ & $n$ & Mean & SD & CS & CK & Mean & SD & CS & CK \\
\midrule
 & Ref. & 1 & 1 & 2 & 9 & 0 & 1 & 0 & 3 \\
\midrule
$0.25$ & 50  & 1.0014 & 0.9896 & 1.5983 & 5.8651 & -0.0011 & 1.0090 & -0.0008 & 2.7994 \\
 & 100 & 1.0009 & 0.9918 & 1.7507 & 6.9224 & -0.0010 & 1.0043 &  0.0048 & 2.8830 \\
 & 200 & 1.0008 & 0.9947 & 1.8415 & 7.5536 & -0.0011 & 1.0021 &  0.0064 & 2.9430 \\
 & 400 & 1.0007 & 0.9968 & 1.9111 & 8.1759 & -0.0009 & 1.0009 &  0.0040 & 2.9654 \\
\midrule
$0.50$ & 50  & 1.0000 & 0.9915 & 1.6072 & 5.9097 &  0.0008 & 1.0091 & -0.0101 & 2.8016 \\
 & 100 & 1.0012 & 0.9925 & 1.7531 & 6.8789 & -0.0013 & 1.0045 &  0.0056 & 2.8867 \\
 & 200 & 1.0018 & 0.9988 & 1.8709 & 7.8367 & -0.0017 & 1.0021 & -0.0061 & 2.9348 \\
 & 400 & 1.0012 & 0.9975 & 1.9202 & 8.2620 & -0.0016 & 1.0009 &  0.0044 & 2.9714 \\
\midrule
$0.75$ & 50  & 1.0030 & 1.0014 & 1.6388 & 6.0228 & -0.0013 & 1.0096 & -0.0325 & 2.8097 \\
 & 100 & 1.0014 & 0.9980 & 1.7728 & 7.0161 & -0.0008 & 1.0046 & -0.0136 & 2.8843 \\
 & 200 & 1.0021 & 0.9996 & 1.8652 & 7.7065 & -0.0017 & 1.0025 & -0.0075 & 2.9304 \\
 & 400 & 1.0020 & 1.0000 & 1.9288 & 8.3314 & -0.0018 & 1.0013 & -0.0041 & 2.9595 \\
\bottomrule
\end{tabular}
\end{table}

\clearpage

\section{Application to real data}\label{sec:5}

In this section, the QLBS regression model is illustrated using meteorological data from the Meteorological Database for Teaching and Research (BDMEP) of the Brazilian National Institute of Meteorology (INMET). The dataset, which was also analyzed by \cite{ocsvils:23} using the LBS regression model, consists of $n = 70$ monthly observations collected at a monitoring station located in Bras\'ilia, Brazil, for the period 2011--2016. The dependent variable $t_i$ is the monthly water evaporation measured in millimeters (mm). The covariates considered in this study are: $x_{1i}$, the actual evapotranspiration (mm); $x_{2i}$, the total insolation (h); $x_{3i}$, the cloudiness (tenths); and $x_{4i}$, the relative humidity (\%). These covariates were identified as statistically significant in the analysis of \cite{ocsvils:23} and are the same as those used here.

\subsection{Descriptive statistics}

\begin{figure}[ht]
\centering
\includegraphics[width=0.9\textwidth]{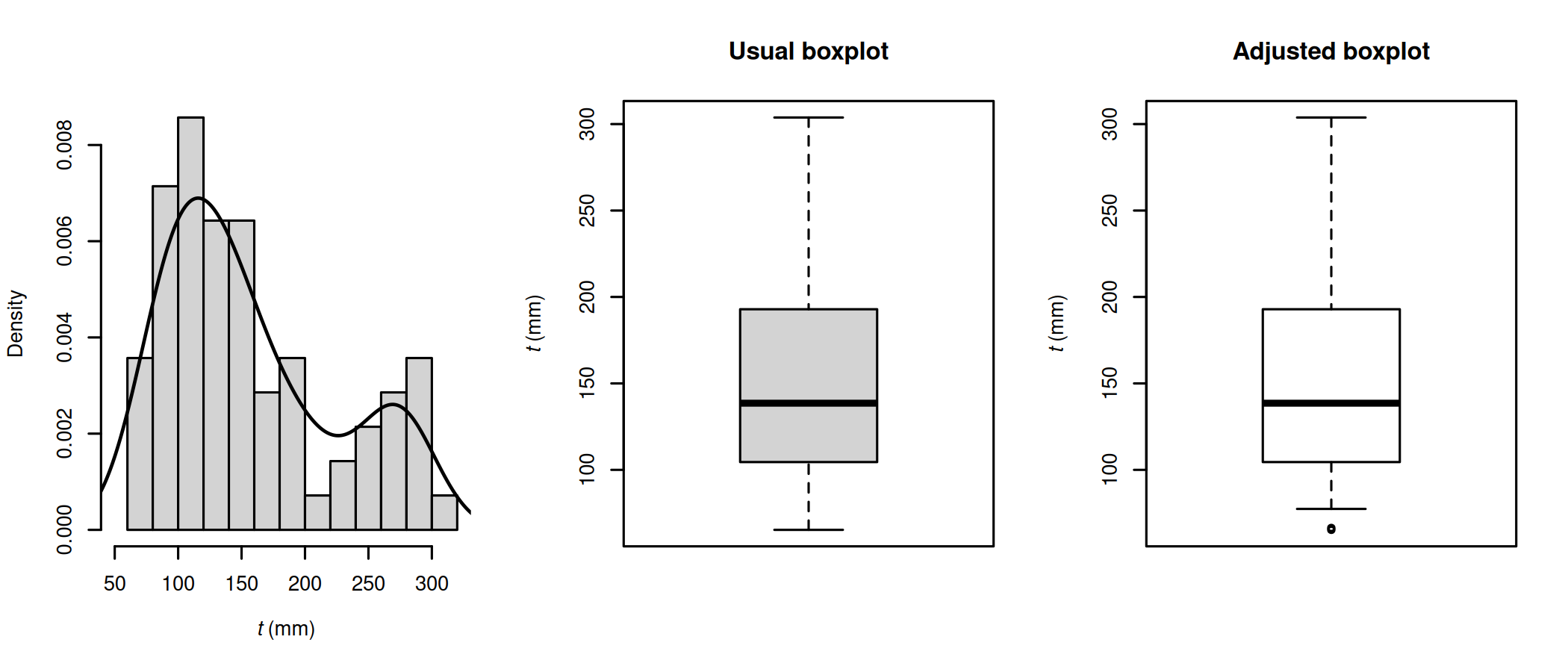}
\caption{Estimated density superimposed on the histogram (left) and usual and adjusted boxplots (center and right) for the water evaporation data.}
\label{fig:density_boxplot}
\end{figure}

Table~\ref{tab:desc} presents the descriptive statistics for the water evaporation response variable. The data exhibit positive skewness and high kurtosis, and the bimodal shape visible in the histogram indicates that the LBS-type models may be particularly appropriate for these data; see Fig.~\ref{fig:density_boxplot}.

\begin{table}[ht]
\centering
\caption{Descriptive statistics for the water evaporation data ($n = 70$).}
\label{tab:desc}
\begin{tabular}{lrrrrrrrrr}
\toprule
Variable & Min & Q$_{0.25}$ & Median & Q$_{0.75}$ & Max & Mean & SD & CV(\%) & CS \\
\midrule
Water evap.\ (mm) & 65.30 & 107.63 & 138.55 & 196.75 & 303.80 & 156.88 & 66.68 & 42.50 & 0.73 \\
\bottomrule
\end{tabular}
\end{table}

\subsection{Estimates and interpretation}

In order to investigate the effect of the meteorological covariates on different parts of the distribution of water evaporation, we fit the QLBS regression model for $\tau = 0.25$, $\tau = 0.50$, and $\tau = 0.75$. Specifically, the QLBS regression model for the $\tau$th conditional quantile is formulated as
\begin{equation}\label{eq:app_model}
\begin{aligned}
\log(Q_{\tau,i}) &= \beta_0 + \beta_1 x_{1i} + \beta_2 x_{2i} + \beta_3 x_{3i} + \beta_4 x_{4i}, \\
\log(\alpha_i)   &= \rho_0 + \rho_1 x_{2i} + \rho_2 x_{3i},
\end{aligned}
\quad i = 1, \ldots, 70.
\end{equation}
The model was fitted using the algorithm described in Section~\ref{sec:3.1}, implemented in R \citep{rmanual:20}. The point estimates, standard errors, and 95\% asymptotic confidence intervals for all three quantile levels are presented in Table~\ref{tab:application}.


\begin{table}[!ht]
\centering
\caption{Point estimates and 95\% asymptotic confidence intervals (ACI) for the QLBS regression model fitted to the water evaporation data.}
\label{tab:application}
\resizebox{\textwidth}{!}{\begin{tabular}{lrrrrrr}
\toprule
 & \multicolumn{2}{c}{$\tau = 0.25$} & \multicolumn{2}{c}{$\tau = 0.50$} & \multicolumn{2}{c}{$\tau = 0.75$} \\
\cmidrule(lr){2-3}\cmidrule(lr){4-5}\cmidrule(lr){6-7}
Parameter & Est.\ (SE) & 95\% ACI & Est.\ (SE) & 95\% ACI & Est.\ (SE) & 95\% ACI \\
\midrule
\multicolumn{7}{l}{\textit{$Q_\tau$ component}} \\
\midrule
Intercept & 6.8317 (0.1756) & (6.4875; 7.1759) & 6.8523 (0.1998) & (6.4607; 7.2440) & 7.0325 (0.2014) & (6.6377; 7.4273) \\
$x_{1i}$ & 0.0015 (0.0004) & (0.0008; 0.0023) & 0.0014 (0.0004) & (0.0006; 0.0022) & 0.0015 (0.0004) & (0.0007; 0.0023) \\
$x_{2i}$ & 0.0009 (0.0005) & (-0.0001; 0.0018) & 0.0008 (0.0005) & (-0.0003; 0.0018) & 0.0003 (0.0005) & (-0.0008; 0.0014) \\
$x_{3i}$ & 0.0391 (0.0156) & (0.0084; 0.0697) & 0.0401 (0.0173) & (0.0061; 0.0741) & 0.0312 (0.0180) & (-0.0041; 0.0665) \\
$x_{4i}$ & -0.0367 (0.0010) & (-0.0386; -0.0347) & -0.0365 (0.0010) & (-0.0384; -0.0345) & -0.0365 (0.0010) & (-0.0384; -0.0346) \\
\midrule
\multicolumn{7}{l}{\textit{$\alpha$ component}} \\
\midrule
Intercept$^\dagger$ & 0.8163 (1.4546) & (-2.0348; 3.6673) & 0.8052 (2.4243) & (-3.9463; 5.5567) & 1.2261 (1.8116) & (-2.3246; 4.7768) \\
$x_{2i}^\dagger$ & -0.0123 (0.0040) & (-0.0201; -0.0044) & -0.0123 (0.0066) & (-0.0253; 0.0007) & -0.0138 (0.0050) & (-0.0236; -0.0039) \\
$x_{3i}^\dagger$ & -0.2192 (0.1150) & (-0.4446; 0.0063) & -0.2153 (0.1874) & (-0.5827; 0.1520) & -0.2383 (0.1399) & (-0.5125; 0.0359) \\
\midrule
Log-lik. & \multicolumn{2}{c}{-244.29} & \multicolumn{2}{c}{-244.18} & \multicolumn{2}{c}{-243.90} \\
AIC & \multicolumn{2}{c}{504.58} & \multicolumn{2}{c}{504.37} & \multicolumn{2}{c}{503.79} \\
BIC & \multicolumn{2}{c}{522.57} & \multicolumn{2}{c}{522.36} & \multicolumn{2}{c}{521.78} \\
\bottomrule
\multicolumn{7}{l}{\small $^\dagger$ $\alpha$ sub-model: Intercept, $x_{2i}$ (insolation), $x_{3i}$ (cloudiness).}
\end{tabular}}
\end{table}

In order to interpret the regression coefficients in Table~\ref{tab:application}, we use the fact that, under the log link, a one-unit increase in covariate $x_{ji}$ multiplies the conditional quantile $Q_{\tau,i}$ by $\exp(\hat{\beta}_j)$, corresponding to a percentage change of $(\exp(\hat{\beta}_j) - 1) \times 100\%$. The 95\% ACIs in Table~\ref{tab:application} are used to assess statistical significance. Below, we present the interpretation of each coefficient.

Actual evapotranspiration ($x_{1i}$, in mm) presents a positive and statistically significant effect on the conditional quantile of water evaporation across all three quantile levels. Specifically, each additional millimeter of actual evapotranspiration is associated with increases of $(\exp(0.0015)-1)\times 100\% \approx 0.15\%$, $(\exp(0.0014)-1)\times 100\% \approx 0.14\%$, and $(\exp(0.0015)-1)\times 100\% \approx 0.15\%$ in the 25th, 50th, and 75th conditional quantiles of water evaporation, respectively, with ACIs entirely above zero at all three levels.

Total insolation ($x_{2i}$, in hours) presents positive point estimates at all three quantile levels---$(\exp(0.0009)-1)\times 100\% \approx 0.09\%$ at $\tau = 0.25$, $0.08\%$ at $\tau = 0.50$, and $0.03\%$ at $\tau = 0.75$. However, the 95\% ACIs include zero at all levels, indicating that the effect of insolation on the conditional quantiles of water evaporation is not statistically significant. Cloudiness ($x_{3i}$, in tenths) presents a positive and statistically significant effect on the lower and median quantiles of water evaporation. Each additional tenth of cloudiness is associated with increases of $(\exp(0.0391)-1)\times 100\% \approx 3.99\%$ at $\tau = 0.25$, $(\exp(0.0401)-1)\times 100\% \approx 4.09\%$ at $\tau = 0.50$, and $(\exp(0.0312)-1)\times 100\% \approx 3.17\%$ at $\tau = 0.75$, whereas the ACI at $\tau = 0.75$ includes zero, suggesting that this effect becomes non-significant for observations in the upper tail of the water evaporation distribution.

Relative humidity ($x_{4i}$, in \%) is the covariate with the largest and most precisely estimated effect on the conditional quantiles of water evaporation. Each additional percentage point of relative humidity is associated with reductions of $(\exp(-0.0367)-1)\times 100\% \approx -3.60\%$, $(\exp(-0.0365)-1)\times 100\% \approx -3.58\%$, and $(\exp(-0.0365)-1)\times 100\% \approx -3.58\%$ in the 25th, 50th, and 75th conditional quantiles, respectively. The 95\% ACIs are entirely below zero at all three quantile levels. We note that this effect is essentially constant across the quantile levels.

The $\alpha$ sub-model describes how the shape parameter $\alpha_i$ of the QLBS distribution, which governs the variability of the response, depends on covariates. Total insolation ($x_{2i}^\dagger$) presents a negative and statistically significant effect on $\log(\alpha_i)$ at $\tau = 0.25$ and $\tau = 0.75$, indicating that higher insolation is associated with lower dispersion in the conditional distribution of water evaporation. Whereas cloudiness ($x_{3i}^\dagger$) consistently presents negative point estimates across all three quantile levels, its ACIs include zero at all levels, providing no statistical evidence of an effect on the shape parameter. The intercept of the $\alpha$ sub-model is not statistically significant at any quantile level.

\subsection{Residual analysis}

\begin{figure}[ht]
\centering
\includegraphics[width=0.90\textwidth]{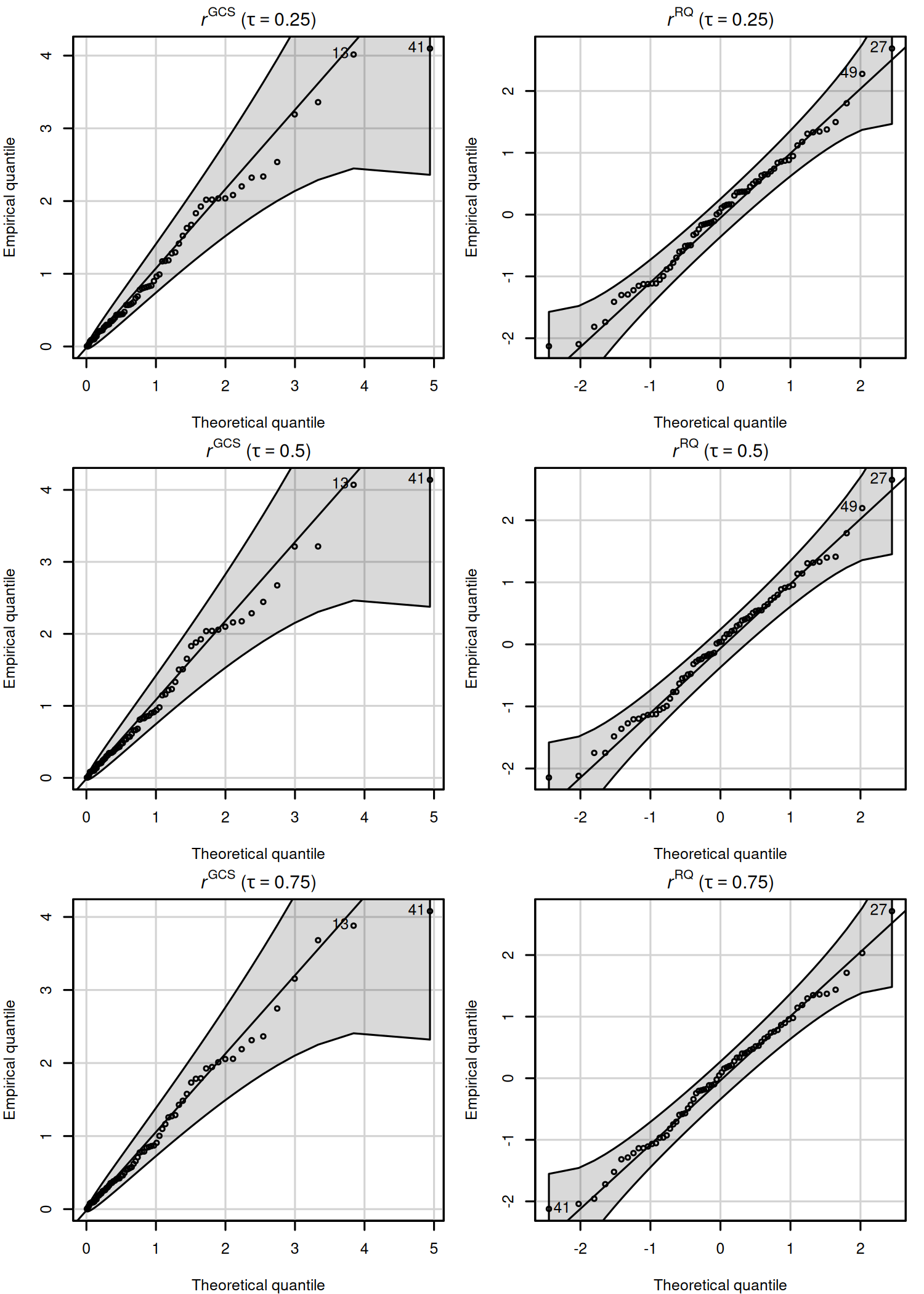}
\caption{QQ plots with simulated envelope for the $r^{\mathrm{GCS}}$ and $r^{\mathrm{RQ}}$ residuals of the QLBS model for $\tau = 0.25$ (top row), $\tau = 0.50$ (middle row), and $\tau = 0.75$ (bottom row), applied to the water evaporation data.}
\label{fig:qqplot_all}
\end{figure}

\begin{figure}[ht]
\centering
\includegraphics[width=0.90\textwidth]{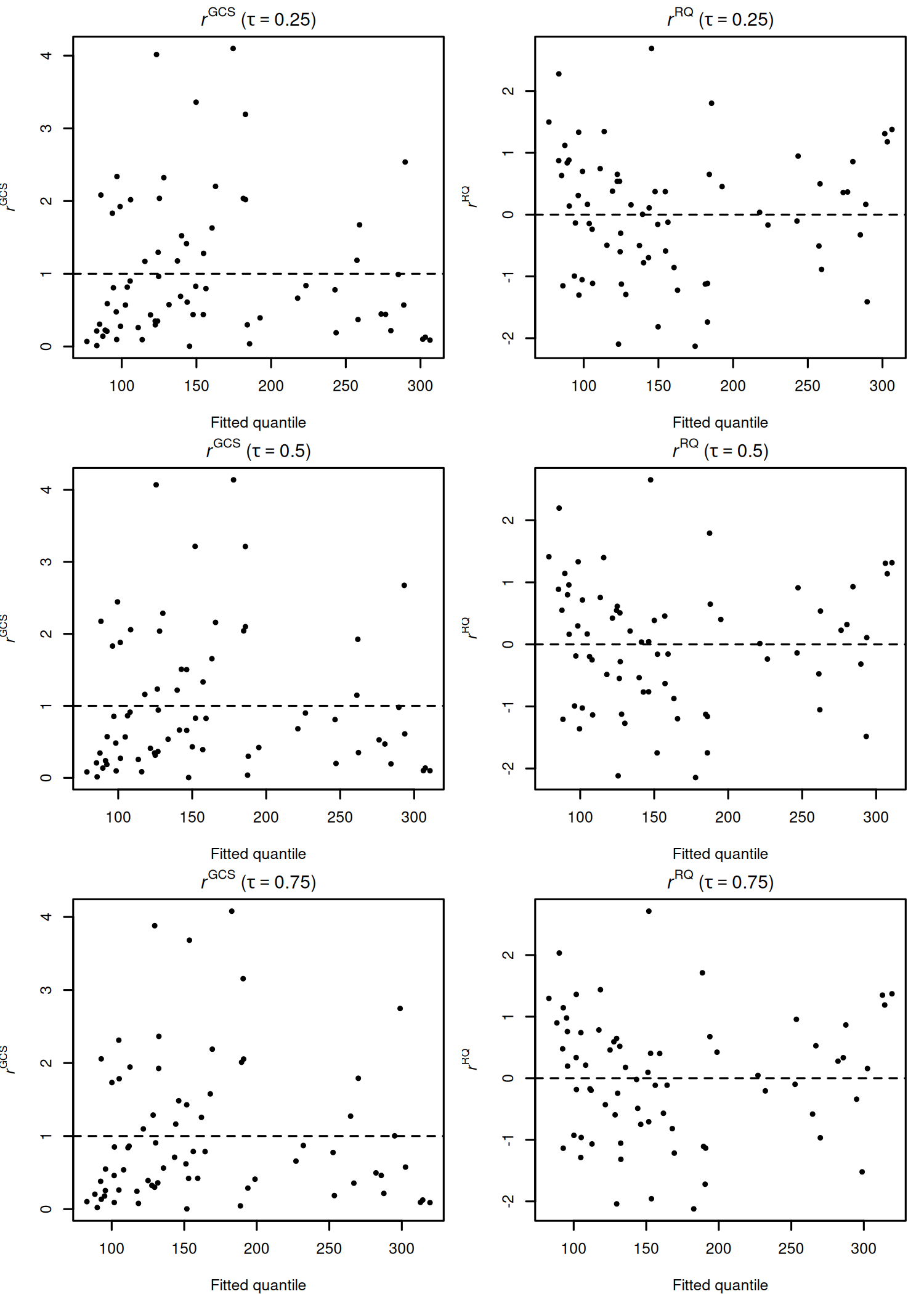}
\caption{Plots of $r^{\mathrm{GCS}}$ and $r^{\mathrm{RQ}}$ residuals against fitted quantiles for the QLBS model for $\tau = 0.25$ (top row), $\tau = 0.50$ (middle row), and $\tau = 0.75$ (bottom row), applied to the water evaporation data.}
\label{fig:res_fitted_all}
\end{figure}

 Figure~\ref{fig:qqplot_all} presents the quantile-quantile (QQ) plots with simulated envelopes for the $r^{\mathrm{GCS}}$ and $r^{\mathrm{RQ}}$ residuals from the QLBS regression model fitted at the three quantile levels $\tau \in \{0.25, 0.50, 0.75\}$. From this figure, we note that the QLBS regression model provides a good fit based on both residuals for $\tau = 0.25$, $0.50$, and $0.75$. Figure~\ref{fig:res_fitted_all} presents the plots of both residuals against the fitted quantiles $\widehat{Q}_{\tau,i}$ for each $\tau$ level, which display random scatter around the respective reference horizontal lines, providing no evidence of systematic departure from the model assumptions at any of the considered quantile levels. Overall, the proposed QLBS regression model fits the meteorological data appropriately across the lower, median, and upper conditional quantiles.

\clearpage

\section{Concluding remarks}\label{sec:6}

Regression models play a central role in statistical analysis, and quantile regression in particular provides a comprehensive and flexible framework for exploring how covariates affect the entire conditional distribution of a response variable. In this paper, we proposed a new quantile regression model based on the length-biased Birnbaum-Saunders distribution. We derived the QLBS distribution by reparameterizing the LBS distribution in terms of its quantile function, thereby allowing the direct modeling of conditional quantiles through a regression structure. We obtained maximum likelihood estimators for the model parameters. Furthermore, we considered two types of diagnostic residuals, the generalized Cox--Snell and randomized quantile residuals. The simulation results show the good performance of the proposed estimators and confidence intervals, with bias, MSE, and coverage probabilities approaching their nominal values as the sample size increases. The proposed QLBS regression model was applied to meteorological data from Brazil, demonstrating its practical utility in describing the effect of meteorological covariates on different conditional quantiles of water evaporation. Extensions of the proposed model to other length-biased distributions, censored data settings, and time series contexts are currently under investigation.

\paragraph{Acknowledgments}
We gratefully acknowledge financial support from CAPES, CNPq and FAP-DF, Brazil.


\section*{Appendix}
\appendix

\section{Derivation of Property (P5)}

Let $Y\sim \mathrm{BS}(\alpha,\beta)$ and define the usual Birnbaum--Saunders transformation
$$
Z=\frac{1}{\alpha}\left(\sqrt{\frac{Y}{\beta}}-\sqrt{\frac{\beta}{Y}}\right),
\qquad\text{so that}\qquad Z\sim N(0,1).
$$
Equivalently, the inverse map is
$$
Y=g(Z)=\frac{\beta}{4}\left(\alpha Z+\sqrt{\alpha^2Z^2+4}\right)^2,
$$
and the change of variables implies
$$
f_Y(g(z))\,g'(z)=\phi(z),\qquad z\in\mathbb{R}.
$$

Now consider the length-biased version $T$ of $Y$, whose density is
$$
f_T(t)=\frac{t\,f_Y(t)}{\mathrm{E}(Y)},\qquad t>0,
$$
with $\mathrm{E}(Y)=\beta(\alpha^2+2)/2$ for the BS model. Using $t=g(z)$ and $dt=g'(z)\,dz$, the induced density of $Z$
under the length-biased law is
$$
f_Z(z)=f_T(g(z))\,g'(z)
      =\frac{g(z)\,f_Y(g(z))}{\mathrm{E}(Y)}\,g'(z)
      =\frac{g(z)}{\mathrm{E}(Y)}\,\phi(z),\qquad z\in\mathbb{R}.
$$

Define the random variable
$$
U=\frac{1}{\alpha^2}\left(\frac{T}{\beta}+\frac{\beta}{T}-2\right).
$$
Since for any $t>0$,
$$
\frac{t}{\beta}+\frac{\beta}{t}-2=\left(\sqrt{\frac{t}{\beta}}-\sqrt{\frac{\beta}{t}}\right)^2,
$$
we obtain the key identity
$$
U=\left[\frac{1}{\alpha}\left(\sqrt{\frac{T}{\beta}}-\sqrt{\frac{\beta}{T}}\right)\right]^2=Z^2.
$$

To find the density of $U=Z^2$, use the standard transformation formula:
for $u>0$,
$$
f_U(u)=\frac{f_Z(\sqrt{u})+f_Z(-\sqrt{u})}{2\sqrt{u}}.
$$
Because $f_Z(z)=\{g(z)/\mathrm{E}(Y)\}\phi(z)$, we need $g(\sqrt{u})+g(-\sqrt{u})$.
From
$$
g(z)=\frac{\beta}{4}\left(\alpha z+\sqrt{\alpha^2z^2+4}\right)^2,
$$
let $s(z)=\sqrt{\alpha^2z^2+4}$ and expand:
$$
g(z)=\frac{\beta}{4}\left(\alpha^2z^2+2\alpha z s(z)+s(z)^2\right)
     =\frac{\beta}{2}\left(\alpha^2z^2+2\right)+\frac{\beta}{2}\alpha z s(z).
$$
Hence,
$$
g(z)+g(-z)=\beta\left(\alpha^2z^2+2\right).
$$
Therefore,
$$
f_U(u)
=\frac{1}{2\sqrt{u}}\frac{g(\sqrt{u})+g(-\sqrt{u})}{\mathrm{E}(Y)}\,\phi(\sqrt{u})
=\frac{1}{2\sqrt{u}}\frac{\beta(\alpha^2u+2)}{\beta(\alpha^2+2)/2}\,\phi(\sqrt{u}),
$$
that is,
$$
f_U(u)=\frac{\alpha^2u+2}{\alpha^2+2}\,\frac{\phi(\sqrt{u})}{\sqrt{u}},\qquad u>0.
$$
Since $\phi(\sqrt{u})=(2\pi)^{-1/2}\exp(-u/2)$, we can rewrite
$$
f_U(u)
=\frac{1}{\alpha^2+2}\,\frac{1}{\sqrt{2\pi}}e^{-u/2}\Big(2u^{-1/2}+\alpha^2u^{1/2}\Big).
$$

Now note that
$$
f_{U_1}(u)=\frac{1}{\sqrt{2\pi}}u^{-1/2}e^{-u/2}\quad\text{is the density of } \Gamma\!\left(\frac12,2\right),
$$
and
$$
f_{U_2}(u)=\frac{1}{\sqrt{2\pi}}u^{1/2}e^{-u/2}\quad\text{is the density of } \Gamma\!\left(\frac32,2\right).
$$
Thus,
$$
f_U(u)=\underbrace{\frac{2}{\alpha^2+2}}_{\pi}\,f_{U_1}(u)
      +\underbrace{\frac{\alpha^2}{\alpha^2+2}}_{1-\pi}\,f_{U_2}(u),
$$
which proves that $U$ is a two-component gamma mixture with
$$
\pi=\frac{2}{\alpha^2+2},\qquad U_1\sim\Gamma\!\left(\frac12,2\right),\qquad U_2\sim\Gamma\!\left(\frac32,2\right).
$$

\end{document}